\documentclass[aps,rmp,onecolumn,superscriptaddress,10pt]{revtex4-2}

\usepackage[T1]{fontenc}
\usepackage{amsmath, amssymb, amsxtra, bm, mathtools, nccmath, tensor, xfrac}
\usepackage{chemmacros}
\chemsetup{modules=all}
\usepackage{siunitx}
\usepackage{graphicx, float, ctable}
\usepackage[inline]{enumitem}
\usepackage{cleveref}
\usepackage[autostyle=false]{csquotes}
\MakeOuterQuote{"}
\usepackage[normalem]{ulem}

\begin{document}

\title{Experimental tests of Lieb--Robinson bounds}
\date{\today}

\author{Marc Cheneau}
\affiliation{Laboratoire Charles Fabry, \\Université Paris-Saclay, Institut d'Optique Graduate School, CNRS, \\91127, Palaiseau, France}


\begin{abstract}
 Judging by the enormous body of work that it has inspired, Elliott Lieb and Derek Robinson's 1972 article on the "Finite Group Velocity of Quantum Spin Systems" can be regarded as a \emph{high-impact paper}, as research accountants say.
 But for more than 30 years this major contribution to quantum physics has remained pretty much confidential.
 Lieb and Robinson's work eventually found a large audience in the years 2000, with the rapid and concomitant development of quantum information theory and experimental platforms enabling the characterisation and manipulation of isolated quantum systems at the single-particle level.
 In this short review article, I will first remind the reader of the central result of Lieb and Robinson's work, namely the existence of a maximum group velocity for the propagation of information in non-relativistic quantum systems.
 I will then review the experiments that most closely relate to this finding, in the sense that they reveal how information propagates in specific---yet "real"---quantum systems.
 Finally, as an outlook, I will attempt to make a connection with the quantum version of the butterfly effect recently studied in chaotic quantum systems.
\end{abstract}

\maketitle

\section{Introduction}

Considering a quantum spin system with finite range interactions and any two operators $A_X$ and $B_Y$, acting on non-overlapping parts of the system $X$ and $Y$, \citet*{Lieb1972} proved that the commutator of $A_X(t)$ and $B_Y(0)$ fulfils a bound of the form:%
\footnote{We do not write the bound in its original form, but rather in a form derived by \citet{Robinson1976} which makes the causal propagation more explicit.}
\begin{equation}
 \label{eq:bound_on_commutator}
 \| [A_X(t), B_Y(0)] \|_\text{op} \leq c \, \|A_X\|_\text{op} \|B_Y\|_\text{op} \, \min(|X|, |Y|) \, \exp\bigg( \frac{v |t| - d}{\xi} \bigg) \; .
\end{equation}
In the left-hand side of this inequality, $A(t) = e^{iHt/\hbar} A \, e^{-iHt/\hbar}$ is the time-evolved operator in the Heisenberg picture and $H$ is the Hamiltonian.
In the right-hand side, $d$ is the distance between $X$ and $Y$, $|X(Y)|$ counts the number of vertices in the subsystem $X (Y)$, and $c$, $v$ and $\xi$ are strictly positive constants which depend on the interaction between the spins and on the lattice geometry.
Importantly, the magnitude of the commutator is measured by the operator norm $\|\cdot\|_\text{op}$, which corresponds to the magnitude of the largest operator eigenvalue.
"The physical content of this statement", as Lieb and Robinson noted, "is that information can propagate in the system only with a finite group velocity".
In a following article, \citet{Robinson1976} elaborated on this result and interpreted the bound in terms of "causal propagation": the action of the operator $B$ at time zero can only affect the operator $A$ at time $t$ if their space-time coordinates lie within the cone $d \leq v|t|$. The velocity $v$ therefore plays the role of an effective speed of light and is referred to as the Lieb--Robinson velocity.

The Lieb--Robinson bound, which applies to the norm of the commutator of two operators, has the advantage of generality since it is independent of the system's state. Unfortunately, it is difficult to link to experimentally or numerically measurable quantities.
This issue was solved simultaneously by \citet*{Bravyi2006} and \citet*{Nachtergaele2006a}, who provided a less general but more operational formulation conveying the same physical content.
Considering an initial state $|\psi\rangle$ with a finite correlation length $\chi$, they showed that the Lieb--Robinson bound implies that it takes a finite time for correlations to build up at distances $d > \chi$.
This is cast into the new bound:
\begin{equation}
 \label{eq:bound_on_correlations}
 |\langle A_X(t) B_Y(t) \rangle_\text{c}| \, \leq c' \, \|A_X\|_\text{op} \|B_Y\|_\text{op} \, (|X| + |Y|) \, \exp\bigg( \frac{2v |t| - d}{\chi'} \bigg) \; ,
\end{equation}
where $\langle A_X(t) B_Y(t) \rangle_\text{c} = \langle A_X(t) B_Y(t) \rangle - \langle A_X(t) \rangle \langle B_Y(t) \rangle $ is the connected correlator, $\chi' = \chi + 2\xi$ and $c'$ is a positive constant.

If a Lieb--Robinson bound establishes the existence of a maximum group velocity for the propagation of information, it should be noted that its \emph{true} value, meaning the smallest $v$ for which \cref{eq:bound_on_commutator} or \labelcref{eq:bound_on_correlations} hold, remains unknown.
Nonetheless, when the low-energy excitations of the system can be effectively described as quasiparticles (or collective excitations), it is tempting to identify the Lieb--Robinson velocity with their maximum group velocity.
\citet*{Calabrese2006} have put flesh on this idea when they showed that entangled pairs of quasiparticles excited by a sudden change in the Hamiltonian (a \emph{quantum quench}) would carry correlations across the system with a finite group velocity given by their dispersion relation, giving rise to a "light-cone effect" \citep[see also][]{Bachmann2016}.
This picture has had a remarkable success for interpreting the dynamics of correlations observed since then in laboratory experiments.
It should however be noted that the Lieb--Robinson bound defines a much more general constraint, not restricted to a low-energy approximation, and will always provide the ultimate limit for the speed at which correlations can propagate.

The prototypical model to which Lieb--Robinson bounds apply is a generic lattice spin model with finite range interactions, but it is natural to seek generalizations to other classes of models.
The first extension one might think of concerns the range of the interaction, as it can be varied in experiments on trapped ions or neutral atoms excited to a Rydberg state.
This issue was briefly addressed by \citet*{Lieb1972}, who explained that the proof of their main theorem could be "refined" to include exponentially or algebraically decaying interactions.
We know today that this prediction was correct, but it took much more work than they may have anticipated to fully understand under which conditions and in which form the locality of the dynamics is preserved when the range of the interaction increases \citep{Chen2019b, Kuwahara2020, Tran2020, Tran2021a}.

The second extension is to go beyond the realm of spin systems and seek a generalization of the Lieb--Robinson bound in quantum boson systems with an infinite-dimension Hilbert space at each lattice vertex.
An example is the Bose--Hubbard model, which can be studied with great precision on different experimental platforms such as ultracold atoms in optical lattices or superconducting circuits.
The theoretical difficulty here lies in the possibility, at least formally, to have an infinite number of bosons clumping on a single site, which makes it impossible to derive a fully general bound.%
\footnote{See for instance the work by \citet{Eisert2009}, where an accelerating signal propagation was engineered using this phenomenon.}
Restricting the study to a finite energy subspace, it was however possible to prove the finite speed of on-site relaxation after a quench \citep{Cramer2008b}, the finite expansion speed of a gas in the vacuum \citep{Schuch2011, Faupin2021a}, the existence of a linear cone bounding the thermal average of the commutator $[A_X(t), B_Y(0)]$ \citep{Yin2022}, a flared cone for the operator norm of $[A_X(t), B_Y(0)]$ \citep{Kuwahara2021}, and, finally, a bound on the speed of macroscopic particle transport \citep{Faupin2022}.

Finally, Lieb--Robinson bounds have also found an application to characterise the dynamical localisation occurring in certain disordered interacting systems \citep{Burrell2007, Hamza2012}. In this case, a zero Lieb--Robinson velocity reflects the absence of propagation of correlations induced by the disorder, or at least indicates that this propagation occurs extremely slowly \citep{Goihl2019}.

I will now move to the central subject of this article, namely the experimental evidences of the finite group velocity at which information propagates in quantum systems.
One should be careful when comparing the observed propagation velocity with the corresponding Lieb--Robinson bound (when it exists), since Lieb--Robinson bounds are intrinsically not tight,%
\footnote{I am aware of one exception to this statement---explicitly mentioned later in this review---, where a quantum state transfer protocol has been shown to saturate a generalized Lieb--Robinson bound in a spin chain with long range interactions \citep{Tran2020, Tran2021b}. This bound however involved the Frobenius norm instead of the operator norm.} %
even if efforts have been made to sharpen the estimate of the Lieb--Robinson velocity.%
\footnote{See for instance \citet{Them2014} or \citet{Wang2020a}, but successive improvements of the bounds are spread throughout the literature.}
Given the huge body of work on the quantum quench dynamics of quantum systems, I have decided to review only those experiments on isolated systems (at least on the time scale of the experiment) and which characterise the propagation of information through the measurement of an equal-time, two-point correlation functions.
This choice can of course be debated, as it leaves aside a number of important works---among which those showing the growth of entanglement entropy over time, a feature that has often been related to the Lieb--Robinson velocity.%
\footnote{The exact link between the propagation of correlations and the growth of entanglement entropy is however not completely elucidated, as exemplified by the case of long-range interacting systems in which distant correlations can build up extremely fast while entanglement entropy grows extremely slowly. See for instance \citet{Lerose2020b} and references therein.}
% The interested reader will find a review of the dynamics of quantum information with a different perspective scope in \citet{LewisSwan2019}.
% In the following, I will primarily sort the works that match my criteria according to the type of model that they map onto.

\section{Experimental evidences for the propagation of information}

\subsection{Lattice boson models}

\begin{figure}
 \centering
 \includegraphics{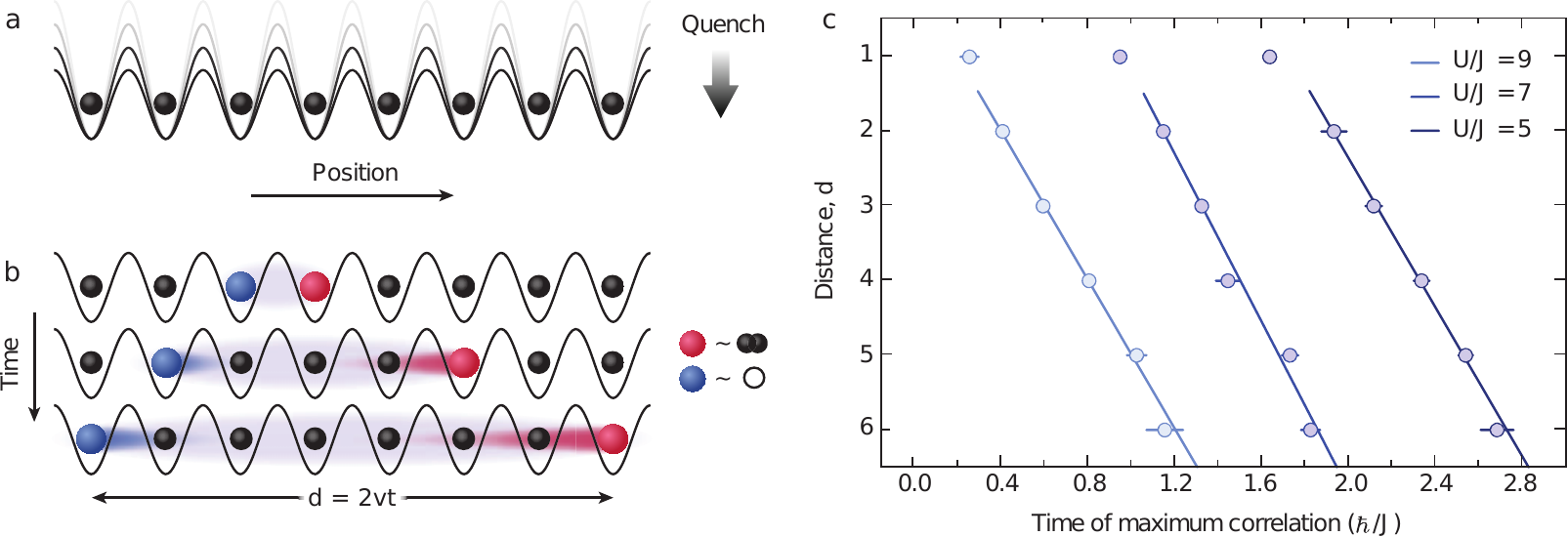}
 \caption{
  (a) \citet{Cheneau2012} loaded an ultracold gas of bosonic atoms in an optical lattice and initialised it in a one-dimensional Mott insulating state with one atom per site.
  (b) At $t=0$, the system is quenched to a lower lattice depth, where the initial state can be viewed as a superposition of counter-propagating quasiparticle pairs. There are two types of quasiparticles, called holons and doublons. A holon (blue ball) is an empty site hybridised with a doubly occupied site by the finite tunnelling amplitude, while a doublon (red ball) is a doubly occupied site hybridised with an empty site. Doublons propagate faster than holons owing to the Bose enhancement of the tunnelling amplitude between a doubly occupied site and a singly occupied site.
  (c) The connected correlations $\langle s_{i} s_{i+d} \rangle$ between the parity of the on-site occupation at sites $i$ and $i+d$---defined as $s_i = \exp(\imath \pi n_i)$---were measured as a function of time. A correlation peak was seen to build up at a time proportional to the distance $d$ (circles). A linear fit through the position of this peak (solid lines) provided an estimate for the propagation velocity $v$.
  The experiment was repeated for different lattice depths, corresponding to different ratios $U/J$, where $U$ is the on-site interaction energy and $J$ is the tunnelling amplitude.
  Figure adapted from \citet{Cheneau2012}.}
 \label{fig:1}
\end{figure}

Our story begins in 2012 when \citet{Cheneau2012} observed for the first time a light-cone-like dynamics in a quantum gas of bosonic atoms loaded in an optical lattice .
The system is accurately described by a one-dimensional Bose--Hubbard model:
\begin{equation}
 H = -J \sum_{\langle i, j \rangle} \big[ a^\dagger_i a^{\vphantom{\dagger}}_j + a^\dagger_j a^{\vphantom{\dagger}}_i\big] + \frac{U}{2} \sum_i n_i(n_i - 1) \; ,
\end{equation}
where $a^{\vphantom{\dagger}}_i \, (a^\dagger_i)$ annihilates (creates) an atom on the lattice site $i$, and $n_i = a^\dagger_i a^{\vphantom{\dagger}}_i$. In this implementation of the Hubbard model with ultracold atoms, the tunnelling amplitude $J$ depends exponentially on the lattice depth, while the on-site interaction energy $U$ is set by the s-wave scattering length between the atoms and is only marginally affected by the lattice depth.
The experiment began by steering a gas of 10 to 20 atoms from the superfluid state to the Mott insulating state with one atom per site by adiabatically increasing the lattice depth to reach $U/J \simeq 40$. This value of $U/J$ is much beyond the critical point of the superfluid-to-insulating transition at $U/J \simeq 3$.
The dynamics was then triggered by quenching the lattice depth to $U/J$ ranging from 5 to 9, that is, closer to the critical point of the model but still in the insulating region (figure~{\ref{fig:1}-a}).
After an evolution time of a fraction of $\hbar/J$ (corresponding to about \SI{1}{\ms} in the experiment), the parity of the on-site occupancy, $s_i = \exp(\imath \pi n_i)$, was measured and the correlator $\langle s_i s_{i+d} \rangle_\text{c}$ was computed by averaging over a large amount of repeated measurements.
As observed in an earlier numerical simulation by \citet*{Lauchli2008}, the propagation of correlations manifested as a peak appearing in the correlation signal at times $t \simeq d / 2v$, with $2v$ varying from about \SI{5}{$J a_{\text{lat}}/\hbar$} at lower values of $U/J$ to about \SI{6}{$J a_{\text{lat}} / \hbar$} at higher values of $U/J$ ($a_\text{lat}$ is the lattice spacing), see figure~{\ref{fig:1}-c}.
The propagation velocity was interpreted in terms of quasiparticle excitations in \citet{Cheneau2012} and \citet{Barmettler2012}: In short, two types of quasiparticles coexist and can be pictured as an empty site weakly hybridised with a doubly-occupied site (\emph{holon}), or as a doubly occupied site weakly hybridised with an empty site (\emph{doublon}, see figure~{\ref{fig:1}-b}). The quench produces entangled doublon-holon pairs with opposite quasimomenta which are responsible for the appearance of density correlations. The velocity at which these correlations propagate is directly given by the sum of the holon and doublon group velocities, and reaches \SI{6}{$J a_{\text{lat}} / \hbar$} in the limit of infinite interactions $U/J \rightarrow \infty$, as observed in the experiment.
In passing, we note that this maximum velocity also matches exactly with a rough upper bound derived by \citet{Cramer2008a} using a "Lieb--Robinson-type argument".

More recently, the Bose--Hubbard model was revisited in a two-dimensional geometry by \citet{Takasu2020}. As in the experiment by \citet{Cheneau2012}, the system was initialised in a deep Mott-insulating state at $U/J \simeq 100$, and quenched by suddenly lowering the lattice depth to $U/J \simeq 20$ (the critical point is at $U/J \simeq 16$ in two dimensions).
After a variable evolution time at constant lattice depth, the gas was released from the lattice and the momentum distribution was measured by imaging the atomic density after a sufficiently long time of flight.
The inverse Fourier transform of this momentum distribution can be expressed as the sum of all phase correlations $\langle a^\dagger_{\bm i} a^{\vphantom{\dagger}}_{\bm j} \rangle$ with fixed $d = \sqrt{|\bm i - \bm j|^2}$ ($\bm i = (i_x, i_y)$ indexes a lattice site in two dimensions).
The evolution of these correlations as a function of distance and time revealed a clear propagation pattern, similar to what was observed by \citet{Cheneau2012} in the one-dimensional geometry.
The extracted velocity for the correlation peak was $v \simeq \text{\SI{13}{$Ja_{\text{lat}}/\hbar$}}$.
Interestingly, this value is significantly larger than the value inferred from quasiparticle propagation, which would be \SI{8.5}{$Ja_{\text{lat}}/\hbar$} for an infinite value of $U/J$.
The authors explain this discrepancy by the difficulty to distinguish the group and the phase velocities in the experimental signal. Interferences between propagating modes indeed produce wiggles in the correlation signal, whose phase drifts linearly with time \citep{Despres2019}. In contrast to the group velocity, the phase velocity is not bounded, but it also conveys no \emph{useful} information.
We may also invoke other reasons for the discrepancy between the observed propagation velocity and the maximum group velocity of quasiparticles
\begin{enumerate*}[label=(\emph{\roman*}), before=\unskip{: }, itemjoin={{; }}]
 \item It was shown by \citet{Cheneau2012} and \citet{Barmettler2012} that the velocity at which the maximum of the correlation function propagates can differ from the maximum group velocity of the quasiparticles at short times (though this effect might be related to the one discussed by \citet{Takasu2020})
 \item The Euclidean distance used to analyse the propagation of correlation in the experiment might not be well suited, especially at short times. Time-dependent variational Monte Carlo simulations indeed showed that the Manhattan metrics was better suited to measure the distance between lattice sites \citep{Carleo2014}.
\end{enumerate*}

\subsection{Bosons in the continuum}

Another bosonic setting was also investigated---this time in a continuous geometry---consisting of a one-dimensional atomic gas in the quasi-condensate regime \citep{Langen2013b}.
The gas was trapped in the magnetic field generated by micro-fabricated wires on a chip.
By applying radio-frequency fields through additional wires, the transverse trapping potential was suddenly transformed into a double well, thereby splitting the initial system into two independent one-dimensional gases parallel to each other.
After an evolution time of up to \SI{10}{\ms}, the trap was switched off and the two subsystems expanded transversally until they overlapped and interfered.
Finally, the matter-wave interference pattern was recorded by absorption imaging and the phase correlation function $\text{Re} \, \left\langle \exp\left(\imath\phi(z,t) - \imath\phi(z+d,t)\right) \right\rangle$ was analysed.
For a given time $t$, the phase correlation was seen to decay exponentially with the distance $d$, like in a thermal state, up to a certain distance $d = 2vt$.
Beyond this front, the correlation was constant, reminiscent of the initial long-range phase coherence.
The velocity at which the correlation front propagated was in qualitative agreement with the speed of sound for a uniform system and in quantitative agreement with the prediction of the Luttinger Liquid theory in a local density approximation \citep{Langen2013b}.
It can thus also be interpreted in terms of quasiparticle propagation.

\subsection{Spin models with short range interactions}

\begin{figure}
 \centering
 \includegraphics{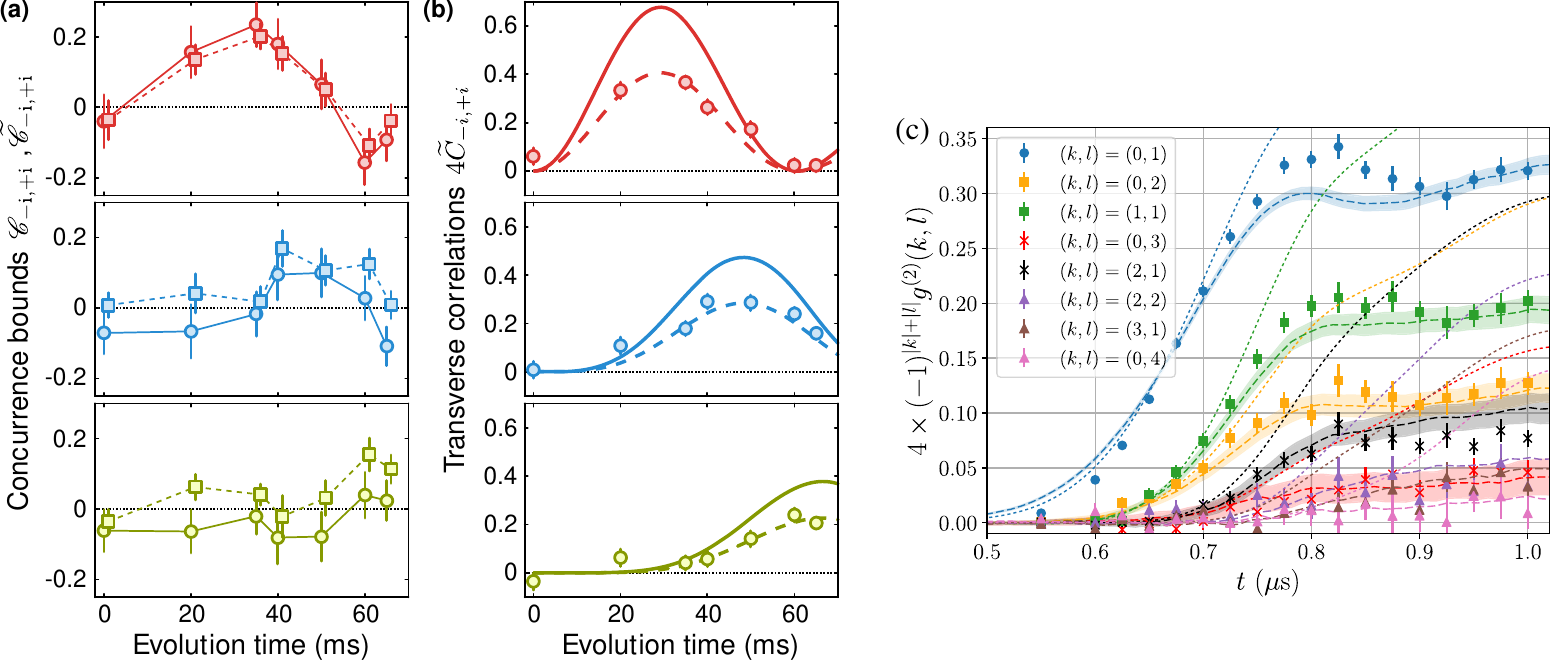}
 \caption{
 (a, b) \citet{Fukuhara2015} prepared a one-dimensional Mott insulating state of neutral atoms with two hyperfine states and used the superexchange coupling between the lattice sites to simulate an Heisenberg XXZ model.
 The system was initialised in a fully polarised state in the $z$-direction and, at $t=0$, a single spin is flipped at the centre of the chain.
 The authors then monitored the dynamics by measuring transverse spin correlations $C_{-i,+i} \propto \langle \sigma^x_{-i} \sigma^x_{+i} \rangle + \langle \sigma^y_{-i} \sigma^y_{+i} \rangle$.
 Combining transverse and longitudinal spin correlations, they also built a lower bound for a quantity known as the concurrence%
 \protect\footnote{The concurrence is a measure of bipartite entanglement which is strictly positive when entanglement is present and grows monotonically with the amount of entanglement \citep{Wootters1998}.}%
 and denoted $\mathcal{C}_{-i,+i}$, which detects the presence of bipartite entanglement between sites $i$ and $j$.
 The tilde versions of these two quantities remove a bias introduced by the presence of holes in the initial state.
 The propagation of correlations manifested by a peak in the correlation/concurrence arising at increasing time for $i = 1$ (red, top row), $i=2$ (blue, middle row) and $i=3$ (green, bottom row).
 In (a), the circles and the squares represent the bare (biased) and the corrected (debiased) experimental values, respectively. The lines simply connect the data points.
 In (b), the circles represent the experimental values of the corrected transverse spin correlation, the solid curves correspond to the analytic predictions in the absence of holes, and the dashed curves are the analytic predictions rescaled by a factor 0.6.
 (c) \citet{Lienhard2018} assembled a $6 \times 6$ array of atoms and excited them optically to a Rydberg state in order to study an antiferromagnetic Ising model with laser-controlled transverse and longitudinal fields.
 The authors adiabatically prepared the system in the ground state at zero fields and then steered it into a state with antiferromagnetic order by slowly varying the laser parameters.
 The authors measured the evolution of antiferromagnetic correlations $g^{(2)}(k, l) \propto \frac{1}{N_{k,l}} \sum_{(\bm i, \bm j)} \langle \sigma^z_{\bm i} \sigma^z_{\bm j} \rangle_\text{c}$, where the sum runs over all pairs of lattice sites $(\bm i, \bm j)$ whose distance is given by $(k \times a_\text{lat}, l \times a_\text{lat})$, $a_\text{lat}$ is the lattice spacing and $N_{k,l}$ is the number of of such pairs of lattice sites.
 The propagation of correlation was apparent in the longer time needed to build these correlations when the distance Manhattan $d = |k| + |l|$ between the sites was increased.
 Figure adapted from \citet{Fukuhara2015} and \citet{Lienhard2018}.
 }
 \label{fig:2}
\end{figure}

Atomic gases were also used to study the propagation of correlations in effective spin lattice systems with nearest-neighbour interactions.
One way to do so is to start again from a Mott insulating state in a deep optical lattice, with one atom per lattice site, and to encode the spin degree of freedom in two atomic hyperfine states, labelled $|\!\downarrow\,\rangle_z$ and $|\!\uparrow\,\rangle_z$.
The superexchange between neighbouring lattice sites then gives rise to an effective coupling between spins, which is cast into a Heisenberg XXZ model
\begin{equation}
 \label{eq:short-range-XXZ}
 H =  \sum_{\langle i,j \rangle}  \left[ \pm J_\perp \left( \sigma^x_i \sigma^x_j + \sigma^y_i \sigma^y_j \right) + J_z \sigma^z_i \sigma^z_j \right] \; ,
\end{equation}
where $\sigma^\gamma_i$ ($\gamma = x, y, z$) are the Pauli matrices, $J_\perp$ and $J_z$ are the transverse and longitudinal superexchange coupling strengths, and the $+$ and $-$ in front of the transverse couplings corresponds, respectively, to fermionic and bosonic atoms.

\citet{Fukuhara2015} realised such a chain of 10 to 20 bosonic atoms and $J_z \simeq J_\perp$.
The system was initialised in the fully polarised state $|\!\downarrow\downarrow\downarrow\ldots\,\rangle_z$ and, at time $t=0$, a local perturbation was applied by flipping a single spin thanks to the combination of a focussed laser beam and a microwave pulse.
The system was then free to evolve for a variable time under the Hamiltonian (\ref{eq:short-range-XXZ}), which reduces to a Heisenberg XY model in the single excitation sector:
\begin{equation}
 H = -J_\perp \sum_{\langle i,j \rangle} \left[ \sigma^+_i \sigma^-_j + \sigma^-_i \sigma^+_j \right] \; ,
\end{equation}
where we have defined $\sigma^\pm = \sigma^x \pm i \sigma^y$.
The system's state was probed by removing the $\downarrow$ atoms and subsequent site-resolved fluorescence imaging of the remaining $\uparrow$ atoms (the removal was necessary due to the lack of spin-sensitivity of the imaging method).
Applying a $\pi/2$ microwave pulse prior to the imaging allowed for a spin measurement in the transverse $xy$ plane.
The computation of the transverse spin correlation $\langle \sigma^x_i \sigma^x_j \rangle + \langle \sigma^y_i \sigma^y_j \rangle$, which is related to the joint probability of detecting two atoms at sites $i$ and $j$ after the $\pi/2$ rotation, then revealed a clear propagation of correlations in space-time (figure~{\ref{fig:2}-b}).
Combining transverse and longitudinal spin correlations to form a lower bound for the concurrence between two lattice sites, the authors could also demonstrate that the build up of correlations between distant sites was concomitant with the appearance of bipartite entanglement (figure~{\ref{fig:2}-a}).

More recently, a new type of experimental platform has been used to study the dynamics of spin lattice systems. It consists of neutral atoms individually loaded in a programmable, two-dimensional array of optical microtraps and excited to Rydberg states.
Introducing again a spin operator to represent the electronic state of the atoms ($\downarrow$ for the ground state and $\uparrow$ for the Rydberg state), one can now map the system onto an Ising model of the form
\begin{equation}
 H = \sum_{\langle \bm i, \bm j\rangle} J_{\bm i \bm j} \sigma^z_{\bm i} \sigma^z_{\bm j} + \frac{\hbar}{2} \sum_{\bm i} \left[ \Omega(t) \sigma^x_{\bm i} - \delta(t) \sigma^z_{\bm i} \right] \; .
\end{equation}
The transverse and longitudinal fields are controlled respectively by the Rabi coupling $\Omega$ and the detuning $\delta$ of the laser exciting the atoms to the Rydberg state.
The interaction $J_{\bm i \bm j}$ arises from the strong van der Waals potential between atoms in a Rydberg state. It can be positive or negative, depending on the Rydberg state, slightly anisotropic because of the shape of the Rydberg orbitals, and mostly couples nearest neighbours owing to the few-micrometer spacing between the optical microtraps.
It should also be noted that this type of systems is only isolated for as long as the Rydberg state does not decay by spontaneous emission or coupling with the surrounding black body radiation.
\citet{Lienhard2018} reported an experiment in which a square lattice of $6\times 6$ sites was initialised in the fully polarised state  $|\!\downarrow\downarrow\downarrow\ldots\,\rangle_z$ and then adiabatically steered into a state with antiferromagnetic order by slowly varying the laser parameters over time.
What's of interest for us is the observation that the joint probability to observe two atoms in the $\uparrow$ state at sites $\bm i$ and $\bm j$ depends on both the (Manhattan) distance between sites and the steering time in a way that clearly indicates a propagation of correlation at a finite velocity of about \SI{6}{$a_{\text{lat}}$ \per \textmu\s} (figure~{\ref{fig:2}-c}).
This velocity was found to be 70 times lower than the estimated the Lieb--Robinson velocity, but in good agreement with the maximum group velocity of the quasiparticle excitations obtained from a linear spin-wave theory.

\subsection{Spin systems with long range interactions}

\begin{figure}
 \centering
 \includegraphics{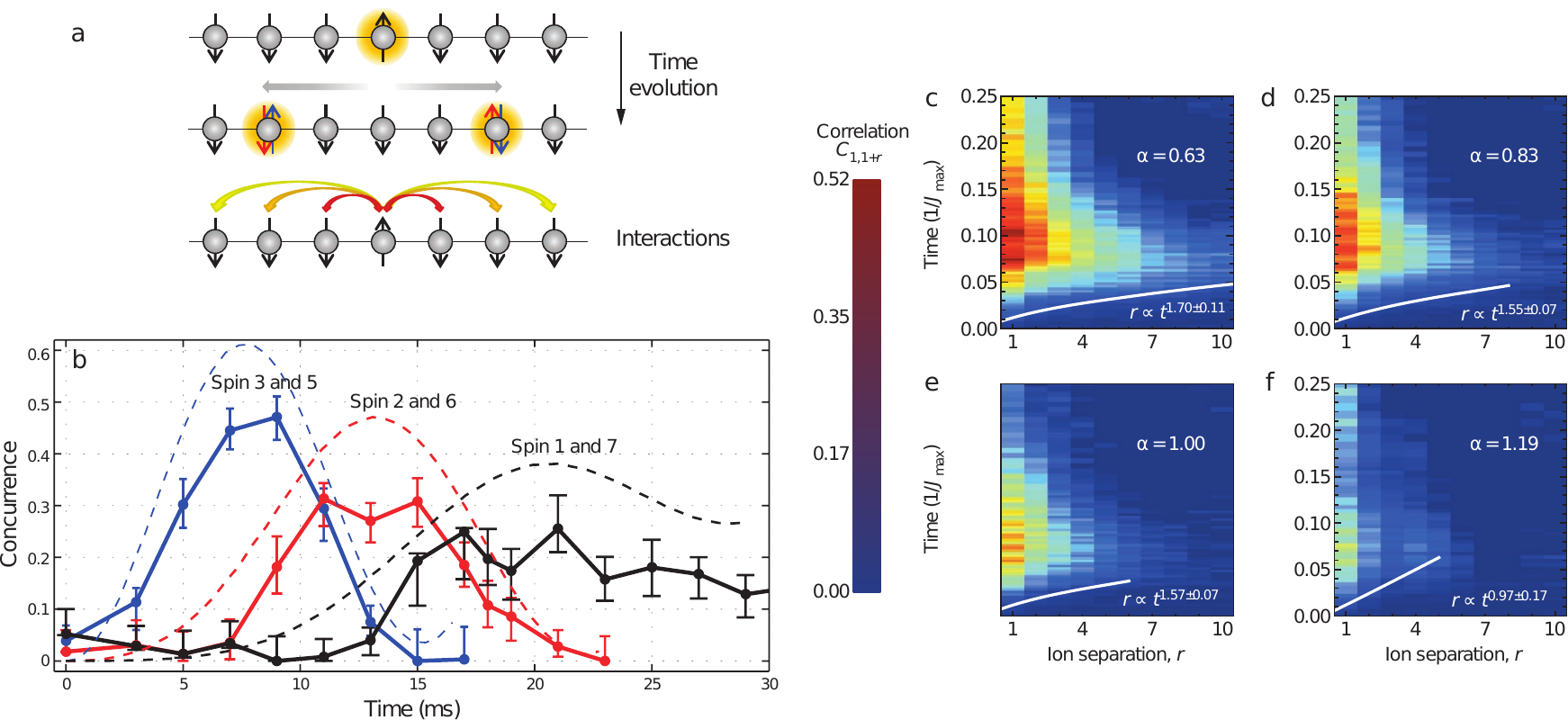}
 \caption{
  Heisenberg XY or Ising models with spin interactions decaying as $1/d^\alpha$ have also been studied using either strings of trapped ions \citep{Richerme2014, Jurcevic2014} or neutral atoms in optical lattices \citep{Zeiher2017}. In the trapped ions systems, the long range interaction was engineered by optically coupling the ions' internal state to collective vibrational modes of the ion chain and the exponent $\alpha$ could be tuned by varying the detuning of the coupling laser. In the neutral atoms system, it was induced by hybridizing the atoms' internal state with a Rydberg state with strong van der Waals interaction, corresponding to $\alpha = 6$.
  (a) Sketch of the experiment by \citet{Jurcevic2014} illustrating the state at the beginning of the dynamics, where the central spin of an initially polarised chain was flipped (top row), and the subsequent propagation of quasiparticle wave-packets, which entangles spin pairs across the system (middle row). The long-range nature of the interactions is also represented (bottom row).
  (b) Performing a full state tomography with $\alpha \approx 1.75$, \citet{Jurcevic2014} were able to measure the time evolution of the concurrence between pairs of spins distributed symmetrically around the central spin (circles). The comparison with a linear spin wave theory (dashed line) shows good qualitative agreement.
  (c) Shows the time evolution of spin correlations $C_{i,i+d}(t) \propto \langle \sigma_i^z(t) \sigma_{i+d}^z(t) \rangle - \langle \sigma_i^z(t) \rangle \langle \sigma_{i+d}^z(t) \rangle$ measured by Richerme \emph{et al.} for different values of the exponent $\alpha$ \citep{Richerme2014}. The data correspond to a situation where the system was prepared in the fully polarised state along $z$ and released in the Ising Hamiltonian with long range interactions. The time is normalised by the value of the spin interaction between nearest neighbours, $J_\text{max}$.
  The solid white lines give a power-law fit to the data.
  Figures adapted from \citet{Richerme2014} and \citet{Jurcevic2014}.
 }
 \label{fig:3}
\end{figure}

Spin systems with algebraically decaying interactions have aroused specific interest because the locality of their dynamics depends on both the exponent characterising the interaction and the dimensionality.
The first experimental platform that has been used for studying the influence of long range interactions consists of strings of 11 to 15 ions held in linear radio-frequency traps \citep{Richerme2014, Jurcevic2014}, and mapping onto a transverse-field Ising model:
\begin{equation}
 \label{eq:long-range-Ising-transverse}
 H = \sum_{\langle i, j\rangle} J_{ij} \sigma^x_i \sigma^x_j + B \sum_i \sigma^z_i \; .
\end{equation}
Here, the internal state of the ions again provides the effective spin-$\sfrac{1}{2}$, and long range spin interaction $J_{ij} \propto 1/|i-j|^\alpha$ and effective magnetic field $B$ both result from the off-resonant coupling of the ions' internal state to their collective transverse modes of motion using laser-driven Raman transition.
The exponent $\alpha$ can be tuned in principle between 0 and 3 by choosing which vibrational modes are addressed by the lasers, but, in practice, only the range $0 < \alpha \lesssim 2$ could been accessed in the experiments.

In the experiment by \citet{Richerme2014}, the system was initialised in the fully polarised state $|\!\downarrow\downarrow\downarrow\ldots\,\rangle_z$ and its time evolution under the Hamiltonian (\ref{eq:long-range-Ising-transverse}) was characterised by the longitudinal spin correlation function $\langle \sigma_i^z(t) \sigma_{i+d}^z(t) \rangle_\text{c}$.
Two regimes were studied:
In the first regime, the transverse field $B$ i set to zero and the dynamics is governed by the long-range Ising model
\begin{equation}
 \label{eq:long-range-Ising}
 H = \sum_{\langle i, j\rangle} J_{ij} \sigma^x_i \sigma^x_j \; .
\end{equation}
This Hamiltonian has a very peculiar structure: because all terms in the sum commute with each other, its excitations are localised and the Lieb--Robinson vanishes.
The system however remains of interest as correlations can still arise between distant spins if they are directly coupled by the interaction, or if they are mutually coupled to a third spin%
\footnote{See the Methods section in \citet{Richerme2014} for a discussion of this effect.}, which, in the experimental setup under discussion here, includes all spins in the chain.
In fact, the authors observed a propagation of the longitudinal spin correlation function under this Ising Hamiltonian, very much resembling what was observed in systems with propagating quasiparticles and a finite Lieb--Robinson velocity (see figure~{\ref{fig:3}-c}).

In the second regime, the transverse field is set to a large value $B \gg \max_{i,j} |J_{ij}|$ and the time evolution of the initial state follows the long-range Heisenberg XY model
\begin{equation}
 \label{eq:long-range-XY}
 H = \sum_{\langle i, j \rangle} J_{ij} \left[ \sigma^+_i \sigma^-_j + \sigma^-_i \sigma^+_j \right] \; .
\end{equation}
In this case also a clear propagation front was observed when plotting the spin correlation function as a function of time and distance.
In contrast to the case of systems with short range interactions, however, the propagation was found to be accelerated, associated with a causal cone of the form $d \propto t^\beta$.
The acceleration was more pronounced for longer-range interactions ($\alpha = 0.63$), and close to zero for shorter-range interactions ($\alpha = 1.19$).

The experiment by \citet{Jurcevic2014}, reported along-side that of \citet{Richerme2014}, brought a complementary perspective on the problem by studying the dynamics after both a global and a local quench in the regime of the Heisenberg XY model.
In the local quench protocol, a tightly focussed laser beam was used to flip a single spin in the initially fully polarised state.
In this single excitation sector, the model is diagonalised by spin waves and the localised excitation can be written as a superposition of quasiparticle wave packets moving in opposite directions (figure~{\ref{fig:3}-a}).
When $\alpha < 2$, owing to the slow decay of the interactions, the maximum group velocity computed from the dispersion relation shows a singularity at a specific value of the quasimomentum.
This analysis of correlations in terms of spin wave propagation was qualitatively confirmed by the measurement of the local magnetisation $\langle \sigma_i^z(t) \rangle$ for $\alpha \gtrsim 1$.
At lower values of $\alpha$, the propagation front of the excitation was less evident and an "almost instant increase in the magnetisation even at large distances" were reported.
Of course, this statement is to be taken carefully given the small system size.
In addition to the dynamics of the local magnetisation, the authors performed a full state tomography for $\alpha \simeq 1.75$ which revealed a spread of entanglement associated with the propagation of the excitation (figure~{\ref{fig:3}-b}).

One last experiment has touched upon the propagation of information in long-range interacting spin systems \citep{Zeiher2017}.
It started similarly to the experiment by \citet{Fukuhara2015}, with a chain of ultracold atoms with two hyperfine states loaded into an optical lattice and prepared in the Mott insulating state with one atom per site.
The main difference with \citet{Fukuhara2015} lies in the fact one of the two hyperfine states (labelled $\uparrow$) was weakly hybridised with a Rydberg state by means of an off-resonant optical coupling.
In a regime where superexchange coupling is effectively suppressed by the great depth of the lattice potential, one then obtains long-range Ising model of the form (\ref{eq:long-range-Ising}), where the interaction $J_{ij}$ is now negative and decays as $1/d^6$.
At the beginning of the experimental sequence of \citet{Zeiher2017}, the system was prepared in the state $|\!\downarrow\downarrow\downarrow\ldots\,\rangle_y$ by rotating the hyperfine state of all atoms with a global $\pi/2$ microwave pulse.
The time evolution of the spin correlations $\langle \sigma^y_i \sigma^y_{i+d} \rangle_\text{c}$ under the Hamiltonian (\ref{eq:long-range-Ising}) was then monitored for $d=1$ and 2.
It was characterised by correlation peak appearing at $t \approx 1/4|J_0|$ when $d=1$ and $t \approx 2/|J_0|$ when $d=2$, with $J_0 = J_{i=j}$.
Although the distance is too short to speak of a propagation of the correlation signal, the observed effect is very reminiscent of that observed by Richerme \emph{et al.} in the Ising limit, and was interpreted by the authors in the same way.

Back in 2014, when the two experiments by \citet{Richerme2014} and \citet{Jurcevic2014} were published, there was no consensus on the type of Lieb--Robinson bound existing in spin systems with power law decaying interactions.
The two experiments showed that the question could be addressed in the lab, but they did not provide a definite answer because of the small system size, which made it difficult to identify the shape of the causal light cone, if any.
The recent theoretical works already mentioned in the introduction \citep{Chen2019b, Kuwahara2020, Tran2020, Tran2021a} have now solved the case by showing that a causal cone always exist, but whose shape depends on $\alpha$ and the lattice dimension $D$: For $\alpha > 2D + 1$, the cone is linear and true Lieb--Robinson bounds exist. For $2D - 1 < \alpha < 2D + 1$, the cone's boundary follows an algebraic curve $t \geq d^{\min (\alpha - 2D - \varepsilon, 1)}$.
Finally, for $\alpha < 2D - 1$, the cone's boundary has a logarithmic shape $t \propto \log d$, which was the first prediction by \citet*{Hastings2006}.

\section{Disordered interacting systems}

If the propagation of correlations seems to characterise the relaxation dynamics of a wide range of physical systems, their localisation%
\footnote{Here I use the word "localisation" in a loose sense encompassing the absence of correlations between distant subsystems at any times, as well as a "glassy" behaviour in which correlations can "leak" towards neighbouring subsystems.}
is thought to be a distinct feature of disordered interacting systems in the so-called many-body localised phase.
This dynamical localisation can take the form of a zero-velocity Lieb--Robinson bound of the form \citep{Sims2013, AbdulRahman2017}:
\begin{equation}
 \| [A_X(t), B_Y(0)] \|_\text{op} \leq c \, \|A_X\|_\text{op} \|Y\|_\text{op} \, \exp\left( - d / \xi \right) \; .
\end{equation}
To the best of my knowledge, such bounds, or variants thereof, have been established so far only for Heisenberg spin chains\citep{AbdulRahman2017, Elgart2018} or coupled harmonic oscillators \citep{AbdulRahman2018}.
Several recent experiments have addressed the dynamical localisation of multi-particle correlations occurring in disordered interacting systems using different proxies:
the entanglement entropy or related measures of many-body entanglement \citep{Smith2016, Xu2018, Lukin2019, Brydges2019, Gong2021b, Chiaro2022},
the evolution of the Hamming distance to the initial state \citep{Smith2016, Gong2021b},
or specially designed measures of the correlation length \citep{Wei2018, Rispoli2019}.
Following the guideline that I have defined for this review, I will only give below some details about the measurements of two-point correlations.

\begin{figure}
 \centering
 \includegraphics{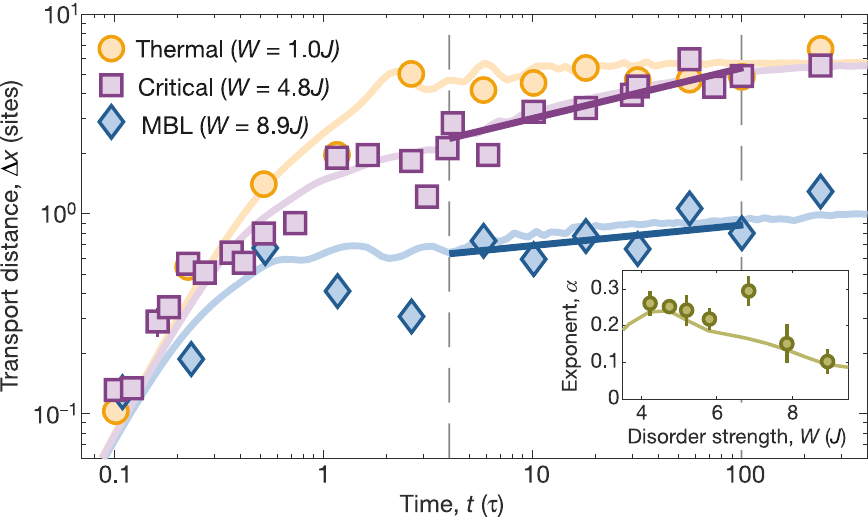}
 \caption{
  \citet{Rispoli2019} report an experiment in which an atomic gas is prepared in the Mott insulating ground state of a one-dimensional optical lattice with quasiperiodic on-site energies, thereby realising an implementation of the Aubry-André model.
  After quenching the tunnelling amplitude to higher values, the transport distance $\Delta x \propto \sum_d d \times \langle n_i n_{i+d} \rangle_\text{c}$ is measured as a function of time, where $\langle n_i n_{i+d} \rangle_\text{c}$ are the connected density correlations between sites separated by a distance $d$.
  After a quick raise at short times, the evolution of transport distance depends on the disorder amplitude $W$:
  At low disorder (yellow circles), $\Delta x$ saturates on a timescale $t / \tau \approx L / 2$, where $\tau = \hbar / J$, $J$ is the tunnelling amplitude and $L$ is the length of the system (up to 12 lattice sites).
  At higher disorder, $\Delta x$ continues to grow slowly following a subdiffusion law $\Delta x \propto t^\alpha$, where $\alpha$ ranges between \num{0.2} at intermediate disorder (purple squares), to \num{0.1} at high disorder (blue diamonds).
  The strongly disordered regime is interpreted as characterising a many-body localised phase, while the intermediate regime is interpreted as a critical region between the thermalising and the localised phases.
  The solid lines accompanying the experimental data are exact numeric calculations.
  Figure adapted from \citet{Rispoli2019}.
 }
 \label{fig:4}
\end{figure}

In the experiment by \citet{Rispoli2019}, a chain of up to 12 bosonic atoms loaded in an optical lattice was prepared in a Mott insulating state, and subject to an additional quasi-periodic optical potential. The Hamiltonian governing the dynamics of this system is known as the Aubry-André Hamiltonian and serves as a model for disordered interacting systems.
It reads
\begin{equation}
 \label{eq:disordered-Bose-Hubbard}
 H = -J \sum_{\langle i, j \rangle} \big[ a^\dagger_i a^{\vphantom{\dagger}}_j + a^\dagger_j a^{\vphantom{\dagger}}_i\big] + \frac{U}{2} \sum_i n_i(n_i - 1) + W \sum_i h_i n_i \; ,
\end{equation}
with $h_i = \cos(2\pi \beta i + \phi)$, $1 / \beta \approx \SI{1.618}{a_{\text{lat}}}$ and $W$ parametrising the amplitude of the disorder distribution.
At time $t=0$, the lattice depth was suddenly reduced to enable the tunnelling of atoms.
After a unitary evolution time, the number of atoms on each site was measured and the connected density correlations $\langle n_i n_{i+d} \rangle_\text{c}$ were computed.
Instead of analysing directly the dynamics of this correlation function as done by \citet{Cheneau2012}, the authors quantified the propagation of correlations through the "transport distance" $\Delta x \propto \sum_d d \times \langle n_i n_{i+d} \rangle_\text{c}$ (see figure~\ref{fig:4}).
At low disorder ($W \sim J$), this quantity was seen to rapidly build up and saturate over a timescale $t \approx \hbar L / 2 J$, where $L$ is the number of sites and $\hbar / J = \SI{4.3}{\ms}$. This saturation, combined with other probes of the growth of many-body correlations, was interpreted as an evidence for the thermalisation of the system.
As the disorder was increased, the saturation of the transport distance was replaced by a slow increase of the form $\Delta x \propto t^\alpha$, with $\alpha$ ranging from \num{0.2} for $W \sim 4-5 J$, to \num{0.1} for $W \sim 10 J$.
This anomalous diffusion was in turn interpreted as the signature of the transition from a thermalising phase to a many-body localised phase with an intermediate critical region.

Most recently, Google's superconducting qubit array was used to study the dynamics of interacting photon excitations in a disordered potential in one and two dimensions \citep{Chiaro2022}. The system is modelled by the same Hamiltonian as in (\ref{eq:disordered-Bose-Hubbard}), except that $h_i \in [-1, 1]$ is now a random variable drawn from a uniform distribution.
The on-site interaction energy in this system is determined by the anharmonicity of the superconducting oscillator that forms the qubit and is fixed at $U / \hbar = \SI{160}{\MHz}$.
The tunnelling amplitude results from the mutual inductive coupling of neighbouring qubits to a coupler loop containing a Josephson junction, such that the inductance, and hence the tunnelling amplitude, can be tuned be applying a magnetic flux through the coupler loop.
Among the different experiments reported by \citet{Chiaro2022}, the one I am specifically interested consisted in initialising a one- or two-dimensional system of 10 to 15 sites in a product state of the form $\left( |0\rangle + |1\rangle \right)_{\bm i_0} \otimes \left( |0\rangle + |1\rangle \right)_{\bm i_1} \otimes |0, 0, \ldots \, \rangle_{\bm j \neq \bm i_0, \bm i_1}$, where two neighbouring sites $\bm i_0$ and $\bm i_1$ are in a superposition of Fock states with 0 or 1 excitation and all other sites are empty.
After a certain evolution time with $J / \hbar = \SI{30}{\MHz}$ and $W = 12 J$, the reduced density matrix on a two-qubit subsystem---consisting of the initial site $\bm i_0$ and a site $\bm i_d$ separated by a linear distance $d = 1, 2, 3$---was measured by state tomography and the so-called entanglement of formation between the two sites was computed%
\footnote{The entanglement of formation of a mixed state $\rho$ is the minimum average
 entanglement of an ensemble of pure states that represents $\rho$ \citep{Wootters1998}. It is directly related to the concept of concurrence of the density matrix that we have already encountered several time in this review.}%
.
In contrast to the von Neumann entropy, which quantifies the entanglement between the subsystem and all its external degrees of freedom, the entanglement of formation discriminates the entanglement between the two qubits inside the subsystem.
Regardless of the dimensionality, the entanglement was seen to propagate towards distant sites with the distance $d$ scaling approximately like $\log t$.

It is not an easy task to make a clear connection between the anomalous diffusion of correlations observed in these two experiments and the existence of a zero-velocity Lieb--Robinson bound.
The first reason is that no zero-velocity Lieb--Robinson bound has been established for the disordered Bose--Hubbard model, although one may argue that the system studied by \citet{Chiaro2022} might alternatively be described as a spin lattice owing to the large on-site interaction energy.
The second reason is that zero-velocity Lieb--Robinson bounds coexist with other signatures of dynamical localisation, such as the logarithmic growth of entanglement entropy,%
\footnote{See \citet{Abanin2019} and references therein.}
which allow for correlations to slowly leak out.
While it seems possible to lift the contradiction between these two point of views by defining different types of Lieb--Robinson bounds \citep{Friesdorf2015}, further work is clearly needed to clarify the relationship between all these concepts.

\section{An alternative measure of locality}

Before moving to the conclusion of this review, I would like to consider recent developments that bare strong similarities with Lieb--Robinson bounds.
\citet{Tran2020} and \citet{Tran2021b} showed that one obtains tighter constraints on the speed of quantum state transfer protocols than that provided by the "usual" Lieb--Robinson bound if one uses the Frobenius norm instead of the operator norm to quantify the non-commutativity of operators.
The Frobenius norm is given by the root mean square of all eigenvalues and therefore provides a more "representative" measure for the norm of an operator, compared to the operator norm used in Lieb--Robinson bounds, which yields a more "conservative" value.
This finding, which suggests that "multiple notions of locality" can coexist \citep{Tran2020}, finds a remarkable echo in recent studies of thermalisation in chaotic quantum systems inspired by a theory of quantum gravity known as the anti-de Sitter/conformal field theory (AdS/CFT) correspondence \citep{Swingle2018}.
The central quantity used to quantify the dynamics in this other context is the expectation value of the squared commutator $[A(t), B(0)]^\dagger [A(t), B(0)]$ over an initially thermal state.
Thermalisation is linked to the spread of the perturbation induced by $A$ at $t=0$, which is itself interpreted in terms of growth of the size of the subsystem over which $A(t)$ acts.
In chaotic systems, the growth of an initially localised operator will eventually lead to large commutators with almost any other operator in the system. This phenomenon is the quantum counterpart of the butterfly effect known in classical chaos \citep{Hosur2016a}.
Interestingly, the operator growth in chaotic spin chains has been found to occur at a finite velocity, called the butterfly velocity \citep{Shenker2014, Roberts2015a}. This naturally reminds us of Lieb and Robinson's effective speed of light, although the two velocities have not been directly related so far.

The expectation value of the squared commutator can also be cast into an out-of-time-order four-point correlator. This is most easily seen when considering Hermitian and unitary operators, for which one can write
\begin{equation}
 \langle [A(t), B(0)]^\dagger [A(t), B(0)] \rangle = 2 - 2 \, \text{Re} \langle A^\dagger(t) B(0)^\dagger A(t) B(0) \rangle \; .
\end{equation}
Such correlators have been measured in a variety of experimental platforms over the past five years: trapped ions \citep{Gaerttner2017, Landsman2019, Joshi2020}, nuclear magnetic resonance on molecules \citep{Li2017} and crystals \citep{Wei2018}, ultracold atoms \citep{Meier2019} and superconducting circuits \citep{Blok2021, Mi2021}.
Two of these experiments reported an observation of the ballistic decay of the out-of-time-order correlator.
\citet{Li2017} studied a four-spin transverse-field Ising model using nuclear magnetic resonance on $\ch{C2F3I}$ molecules dissolved in a solvent.
The forward and backward time evolution was encoded in a sequence of single- and two-spin operations performed by radio-frequency pulses and the operators $A$ and $B$ were chosen to be Pauli operators acting on the individual spins $i=1$ and $j > i$.
The expectation value of the out-of-time-order correlator in a fictitious infinite temperature state was measured through the expectation value of the local Pauli operator acting on the first spin.
It was then clearly observed that the time at which the correlator decayed was proportional to the distance $d = |i - j|$, thereby providing a direct measure of the butterfly velocity in this system.
Most recently, a quantum circuit with random components was programmed on Google's superconducting qubit array \citep{Mi2021}.
Here again, the operators $A$ and $B$ were taken to be local Pauli operators acting on different spins.
The expectation value of the out-of-time-order correlator was measured using an interferometric protocol that maps it to the expectation value of the spin of an ancillary qubit.
Two types of circuits were implemented: a chain of 21 qubits and a two-dimensional array of 53 qubits. In both cases, the decay of the out-of-time-order correlator exhibited a clear ballistic signature.

\section{Conclusion}

With this short review, I have tried to draw a complete picture of the experimental evidences for the propagation of information in the form of two-point correlations with a finite group velocity in isolated quantum systems.
Such evidences have been found in a wide variety of experimental systems, which confirms the idea that the locality of the dynamics is a widely spread feature of quantum systems.
In most experiments, the propagation of correlations was interpreted with a quasiparticle picture and the propagation velocity was found in fair agreement with the maximum group velocity computed from the dispersion relation.
This may give the impression that the Lieb--Robinson velocity and the maximum group velocity of quasiparticles are two sides of the same coin.
In reality, the connection between the two concepts is far from obvious and experiments have been of little help so far to clarify this point because they were performed on small systems.
This is an issue for several reasons.
First, interferences related to the phase velocity can strongly modulate the correlation signals at short space-time distances, making it difficult to identify the propagation due to the group velocity.
Second, the different scalings of the propagation distance with time, depending on the range of the interactions or the presence of disorder, can only be distinguished after sufficiently long evolution times.
And, finally, it would be most interesting to see how the finite lifetime of the quasiparticles affect the propagation of correlations in the experiments, which, again, requires long evolution times.

In order to address these issues, it is necessary to push the experiments towards larger system sizes.
Systems of several hundreds of spins have recently been engineered in arrays of Rydberg atoms \citep{Scholl2021}, but the finite lifetime of Rydberg states will ultimately cap the accessible unitary evolution times.
The newer generation of optical lattice experiments with single-particle imaging resolution, as well as the constant progress towards larger superconducting qubits arrays, will also help in this regard.
Another approach would be to probe the dynamics in critical systems were the lifetime of quasiparticles is very short. An example of such system which is already accessible experimentally is the two-dimensional Bose--Hubbard model \citep{WitczakKrempa2014} in the vicinity of the superfluid-to-insulator transition.
I don't know if the observation reported by \citet{Takasu2020} that phase correlations propagate faster than the quasiparticle prediction can be related to the criticality of the Bose--Hubbard model, but it would definitely be interesting to devise specific experiments in order to test the limits of the quasiparticle picture.

Finally, I have opened the review to an alternative measure of locality in terms of operator growth, which manifests in chaotic spin systems by a light-cone-like structure in the decay of out-of-time-order correlators.
The similarity between the two approaches is striking and I am eager to see if a formal connection can be made between the two, beyond the observation that they provide similar bounds on locality \citep{Colmenarez2020}.
For the time being, the focus of experimentalists seems to have switched from Lieb--Robinson bounds to the butterfly effect, and several measurements of out-of-time-order correlations have been reported lately.
Such correlators are generally difficult to measure directly in AMO systems because of the need to reverse the sign of the time evolution, but the proposal by Vermersch \emph{et al.} to rely on randomised measurements of simpler correlators is a promising route to explore \citep{Vermersch2019, Joshi2020}.
It is also interesting to note that the rapidly developing programmable digital quantum simulators, containing today tens of superconducting circuits, has also found there an ideal test-bed for their computing power.
My hope now is that experimental efforts will be sustained after the premieres in order to truly contribute to the understanding of this fascinating topic.

\bibliography{../experimental_tests_of_lieb-robinson_bounds.bib}

\end{document}